# Dynamics of COVID-19 mitigation inefficiency in Brazil

Borko D. Stosic[1*]

[1] Universidade Federal Rural de Pernambuco, Departamento de Estatística e Informática, Rua Dom Manoel de Medeiros s/n, Dois Irmãos, 52171-900 Recife/PE, Brazil

*Corresponding author. Email: borko.stosic@ufrpe.br

**Abstract:**

In this work Data Envelopment Analysis (DEA) is employed in thirty-day windows to quantify temporal evolution of relative pandemic mitigation inefficiency of Brazilian municipalities. For each thirty-day window the results of inefficiency scores of over five thousand Brazilian municipalities are displayed on maps, to address the spatial distribution of the corresponding values. This phenomenological spatiotemporal approach reveals location of the hotspots, in terms of relative pandemic inefficiency of the municipalities, at different points in time along the pandemic. It is expected that the current approach may subsidize decision making through comparative analysis of previous practice, and thus contribute to the future pandemic mitigation efforts.

**Introduction**

A year into the COVID-19 pandemic, Brazil, as many other countries, is currently experiencing a severe second wave of the disease, currently being described as the world's epicenter of the pandemic. Coronavirus disease 2019 (COVID-19), a highly contagious disease caused by severe acute respiratory syndrome coronavirus 2 (SARS-CoV-2), was declared officially a pandemic by World Health Organization on March 11, 2020 [1]. Emerging in China, and initially being mostly contained in that country, the virus has been spreading worldwide ever since. This dispersion has invoked stringent protective measures against the spread in most countries in the world, ranging from self-isolation and mandatory quarantine, to curfews and travel restrictions.

The current situation remains precarious: different countries have been implementing different measures to mitigate the propagation of the virus ever since the pandemic start, but the outcome of the effectiveness of these diverse measures has yet to be quantified. In Brazil, a continental size country with over two hundred million inhabitants and over five thousand municipalities, a wide range of mitigation measures has been implemented by the local governments, at different time frames. The overall result of these measures is currently often described in the news as disastrous, with an overloaded health system and a large number of deaths of people waiting for a bed in intensive care units. On the other hand, Brazil has a very high organizational vaccination capacity through the highly organized public health system (Sistema Unico de Saude - SUS), but the world's capacity of vaccine production has turned insufficient to provide the number of vaccine doses necessarily to reach collective heard immunity. This is the current problem in most countries, indicating that a prolonged difficult period still lies ahead. Consequently, it appears paramount to expand the analysis of the available data in different directions, in the attempt of providing complementary insight into the phenomenon, in order to aid the policymakers in their decisions.

To this end, Data Envelopment Analysis (DEA) is employed in this work to quantify mitigation inefficiency of 5442 Brazilian municipalities along the year of the pandemic, in a spatially explicit manner. It is expected that the results of this work may aid a governor of a state, or a major of a city, to compare their own measures, taken at different times, to those of neighboring states or cities, and draw conclusions as to the efficiency of their own disease mitigation measures.

**Data and methodology**

*Data*

The database used in this study [2], available at https://github.com/wcota/covid19br, contains data on 5570 Brazilian municipalities, including population size, location (latitude and longitude), and daily numbers of confirmed cases and deaths, since February 25, 2020 (the data used in this work are up to April 12, 2021). It was found that 128 (very small) municipalities have no reported deaths along the period under study, and as the subsequent analysis is sensitive to outliers they were removed from this study, which was performed with the remaining 5442 municipalities.

*Data Envelopment Analysis*

In what follows Data Envelopment Analysis (DEA) method is used to address mitigation inefficiency of the Brazilian municipalities. DEA was introduced in econometrics, it represents a powerful, non-parametric method for measurement of relative production efficiency of a given set of comparable profit organizations. The concept of DEA was originally introduced by Farrell [3], and subsequently formulated as a Linear Programming based technique by Charnes, Cooper and Rhodes [4] (the reader is referred to a recent textbook by Charnes et al. [5] for a comprehensive coverage of the subject). The method is achieved by considering empirical efficiency frontiers, spanned by such organizations (so called Decision Making Units – DMUs), which use minimal input to produce maximal output within the observed sample. While DEA was originally developed for comparative efficiency measurement of economic entities within a strict profit maximization concept (banks, industries, firms, etc.), it has been subsequently applied to various multiple input and output situations, where an input is understood as a quantity that should be minimized, and an output as a quantity that should be maximized, in order to reach the so called "efficiency frontier".

The DEA method proceeds as follows. For a set of $K$ DMUs, $j = 1, \dots, K$, with $N$ (positive) inputs $x_{jn}$, $n = 1, \dots, N$ and $M$ (positive) outputs $y_{jm}$, $m = 1, \dots, M$ the idea of the DEA method lies in maximizing for each of the DMUs the ratio $h_k$ of the linear combination of all the outputs weighted by positive coefficients $u_m$ and the linear combination of the inputs weighted by positive coefficients $v_n$

$$max_{u,v} h_k = \frac{\sum_{m=1}^{M} u_m y_{km}}{\sum_{n=1}^{N} v_n x_{kn}} \tag{1}$$

subject to restrictions

$$\frac{\sum_{m=1}^{M} u_m y_{jm}}{\sum_{n=1}^{N} v_n x_{jn}} \leq 1 \qquad j = 1, \dots, k, \dots, K \tag{2}$$

$$u_m, v_n \geq 0 \qquad m = 1, \dots, M \; ; \quad n = 1, \dots, N \tag{3}$$

Effectively, this represents scaling of all of the input and output variables which maximizes the output versus input ratio of the considered DMU $k$, while making sure (by imposing restriction (2)) that similar ratios (comparative efficiencies) for all of the DMUs remain bounded by unity. The conditions (3) guarantee the required positivity of the input and output variables after scaling. It is evident that any solution to relations (1-3), represented by sets of u's and v's, is invariant to uniform scaling, achieved by multiplying all the u's and the v's by the same (arbitrary, positive) constant. As this generates an infinite set of solutions, the usual approach to overcome this problem and remove the solution degeneracy is to impose the additional constraints, either $\sum_{n=1}^{N} v_n x_{jn} = 1$ (so called input orientation), or $\sum_{m=1}^{M} u_m y_m = 1$ (output orientation). In the case of input orientation, the procedure is now reduced to determining

$$max_{u,v} h_k = \sum_{m=1}^{M} u_m y_{km} \qquad (4)$$

subject to

$$\sum_{n=1}^{N} v_n x_{jn} = 1 \qquad (5)$$

$$\sum_{m=1}^{M} u_m y_{jm} - \sum_{n=1}^{N} v_n x_{jn} \leq 0 \qquad j = 1, \ldots, k, \ldots, K \qquad (6)$$

$$u_m, v_n \geq 0 \qquad m = 1, \ldots, M \; ; \quad n = 1, \ldots, N \qquad (7)$$

which is called input oriented DEA "multipler formulation". Finally, it turns out that every linear programming problem has two equivalent formulations, and that the dual formulation (so called "envelopment form") is numerically more advantageous (as there are more DMUs than inputs and outputs), so the common input oriented DEA formulation found in the literature is given by

$$\theta_k \equiv min_{\theta,\lambda}(\theta) \qquad (8)$$

subject to

$$\theta x_{kn} \geq \sum_{j=1}^{K} \lambda_j x_{jn} \qquad n = 1, \ldots, N \qquad (9)$$

$$y_{km} \leq \sum_{j=1}^{K} \lambda_j y_{jm} \qquad m = 1, \ldots, M \qquad (10)$$

$$\theta, \lambda_j \geq 0 \qquad j = 1, \ldots, K \qquad (11)$$

Here $\lambda_j$ are positive coefficients to be adjusted (for each DMU $k$) in order to minimize $\theta_k$ (instead of scale factors $u_m$ and $v_n$ in the multiplier formulation (4-7)). For each DMU with efficiency below unity, this input oriented DEA formulation indicates as a goal the reduction of inputs necessary for reaching the efficiency frontier.

In the output oriented DEA version, by normalizing the linear combination of outputs, relations (1-3) are transformed to yield the multiplier form

$$min_{u,v} g_k = \sum_{n=1}^{N} v_n x_{kn} \qquad (12)$$

subject to

$$\sum_{m=1}^{M} u_m y_m = 1 \qquad (13)$$

$$\sum_{m=1}^{M} u_m y_{jm} - \sum_{n=1}^{N} v_n x_{jn} \leq 0 \qquad j = 1, \ldots, k, \ldots, K \qquad (14)$$

$$u_m, v_n \geq 0 \qquad m = 1, \ldots, M \ ; \quad n = 1, \ldots, N \qquad (15)$$

Finally, the linear programming problem dual to relations (12-15) yields the envelopment form

$$\omega_k = max_{\omega, \lambda}(\omega)) \qquad (16)$$

subject to conditions

$$\omega y_{km} \leq \sum_{j=1}^{K} \lambda_j y_{jm} \qquad m = 1, \ldots, M \qquad (17)$$

$$x_{kn} \geq \sum_{j=1}^{K} \lambda_j x_{jn} \qquad n = 1, \ldots, N \qquad (18)$$

$$\omega, \lambda_j \geq 0 \qquad j = 1, \ldots, K \qquad (19)$$

and the efficiency of the $k$-th DMU is given by $\theta_k = 1/\omega_k$.

Both the above versions of DEA are said to have constant returns to scale (CRS), referring to the assumption that outputs can grow indefinitely with input increase, and approach zero as the inputs go to zero. In order to account for possible scale effects for very small and very large DMUs, Banker, Charnes and Cooper [6] have proposed the so called variable returns to scale (VRS) version of DEA, which is implemented by adding a single additional constraint

$$\sum_{j=1}^{K} \lambda_j = 1$$

to the envelopment forms (8-11) and (16-19). All the above versions of DEA represent typical linear programming problems, and are solved numerically.

To examine the "efficiency frontier", in the current context of COVID 19 the numbers of confirmed cases and deaths can be taken as an input, and municipality population as output, in which case the so called input oriented DEA is adequate, which considers the reduction of inputs for which the observed DMU reaches the efficiency frontier. On the other hand, if the main interest is to identify the most inefficient municipalities (the "inefficiency frontier"), population can be taken as input, and the numbers of confirmed cases and deaths taken as outputs: if one of two particular municipalities, with the same population size, has more confirmed cases, and/or more deaths, it is considered to exhibit a higher mitigation inefficiency than the other. In this case the so called output oriented DEA is implemented, which considers the increase of the quantity of outputs (here confirmed cases and deaths) at which the observed DMU reaches the "inefficiency frontier".

*Inverse Distance Weighting*

The spatial distribution of the values is obtained in this work by interpolation using standard isotropic Inverse Distance Weighting (IDW) technique, introduced by Shepard [7]. Interpolated value of the quantity $u(x)$ at point $x$ is found from the known values $u(x_i)$ at the $K$ neighboring data points $x_i, i = 1, \ldots, K$, using

$$u(x) = \frac{\sum_{i=1}^{K} w_i u(x_i)}{\sum_{i=1}^{K} w_i} \tag{7}$$

where

$$w_i = \frac{1}{d(x, x_i)^p} \tag{8}$$

and $d(x, x_i) \neq 0$ is the distance between the points $x$ and $x_i$. The actual value of the parameter $p$ is taken to be $p = 2$ throughout this work, as originally proposed by Shepard [7].

**Results and discussion**

The descriptive statistics of the relevant variables: municipality population size, number of confirmed cases over the period of study, confirmed percentage (number of confirmed cases per 100 inhabitants), number of deaths, deaths per 100000 inhabitants, death rate (number of deaths per hundred confirmed cases), and the DEA scores for the efficiency and the inefficiency versions, in both CRS and VRS versions, are presented in Tab. 1.

**Table 1.** The descriptive statistics of relevant variables and DEA scores (see text for more details).

|  | Minimum | 1st quartile | Median | Mean | 3rd quartile | Maximum | Standard deviation |
|---|---|---|---|---|---|---|---|
| Population | 837 | 5638 | 11957 | 38466 | 25764 | 12252023 | 223990 |
| Confirmed cases | 10 | 263 | 580 | 2467 | 1506 | 663212 | 13387 |
| Confirmed percentage | 0.151 | 3.636 | 5.445 | 6.026 | 7.721 | 33.316 | 3.338 |
| Deaths | 1 | 5 | 12 | 64.90 | 30 | 24282 | 518.6 |
| Deaths per 100k | 4.495 | 69.990 | 111.198 | 124.019 | 163.040 | 647.848 | 73.583 |
| Death rate (%) | 0.120 | 1.408 | 2.046 | 2.371 | 2.894 | 23.077 | 1.562 |
| DEA efficiency CRS | 0.0123 | 0.045 | 0.066 | 0.087 | 0.101 | 1.0 | 0.075 |
| DEA efficiency VRS | 0.0200 | 0.093 | 0.167 | 0.239 | 0.294 | 1.0 | 0.228 |
| DEA inefficiency CRS | 0.0129 | 0.160 | 0.233 | 0.254 | 0.325 | 1.0 | 0.129 |
| DEA inefficiency VRS | 0.0170 | 0.184 | 0.279 | 0.311 | 0.403 | 1.0 | 0.170 |

The range of the data of four orders of magnitude indicates use of a logarithmic scale for adequate display of the histograms of population, number of confirmed cases, and number of deaths, as presented in Fig 1.

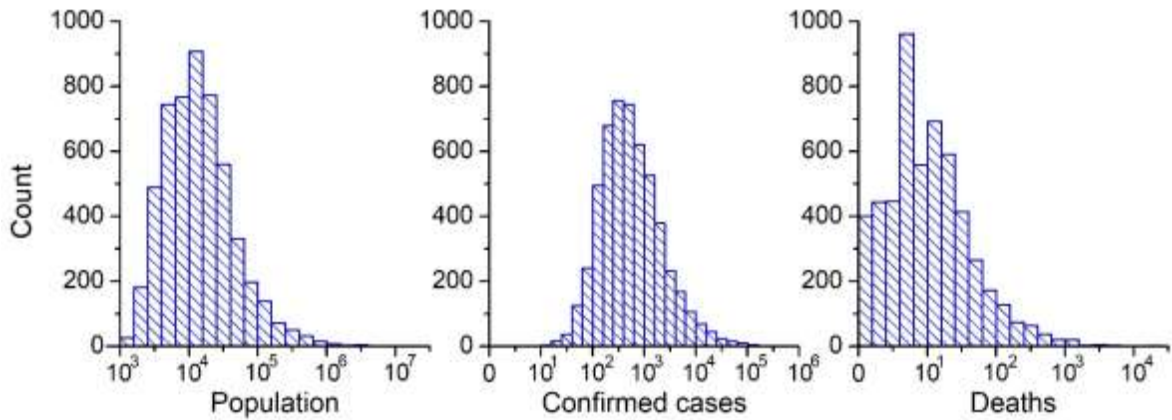

**Fig. 1.** Histograms of the population count, total confirmed cases, and deaths in the period of observation, on logarithmic scale.

It is seen from Tab. 1, and Fig. 1 that population, confirmed cases and number of deaths span several orders of magnitude, suggestive of power law behavior. In fact, the validity of Zipf's law [8,9] (also called rank-size rule) has already been shown for Brazilian cities [10], and for COVID-19 cases for a number of countries [11]. Zipf's law holds for a given dataset if for the ordered values $Z_1 > Z_2 > \cdots > Z_r > \cdots > Z_N$ the plot of $ln(Z_r)$ versus $ln(r)$ is linear. In the current case, all three quantities follow the Zipf's law for a wide range of values, with a rather similar exponent (slope on the log-log plot), as can be seen in Fig. 2.

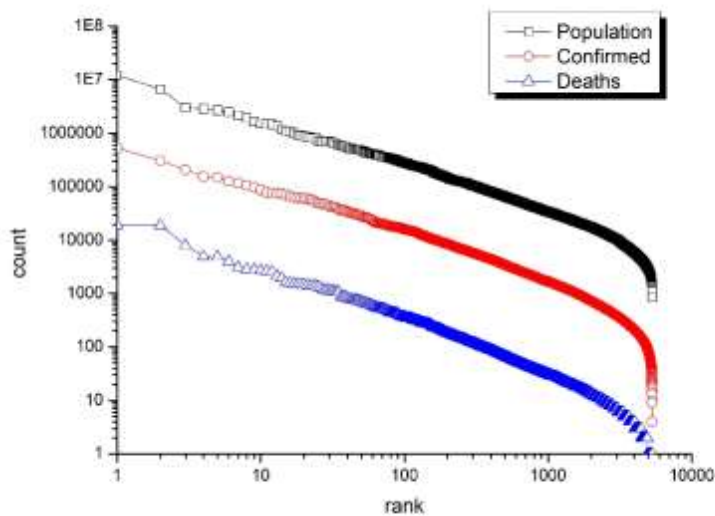

**Fig. 2.** Value versus rank of the population count, total confirmed cases and deaths, for 5442 Brazilian municipalities, in the period of observation, on double logarithmic scale. Linear behavior in a wide range confirms validity of Zipf's law.

The spatial distribution of population, confirmed cases and deaths over the study period is shown in Fig. 3 (again on logarithmic scale), together with the death rate (number of deaths divided by the number of confirmed cases, on linear scale). It is seen that the spatial distribution of confirmed cases and deaths closely matches the municipality population size distribution, while the death rate spatial distribution reveals a rather different pattern, that should be more scrutinized by local governments of municipalities shown in red on Fig. 3d.

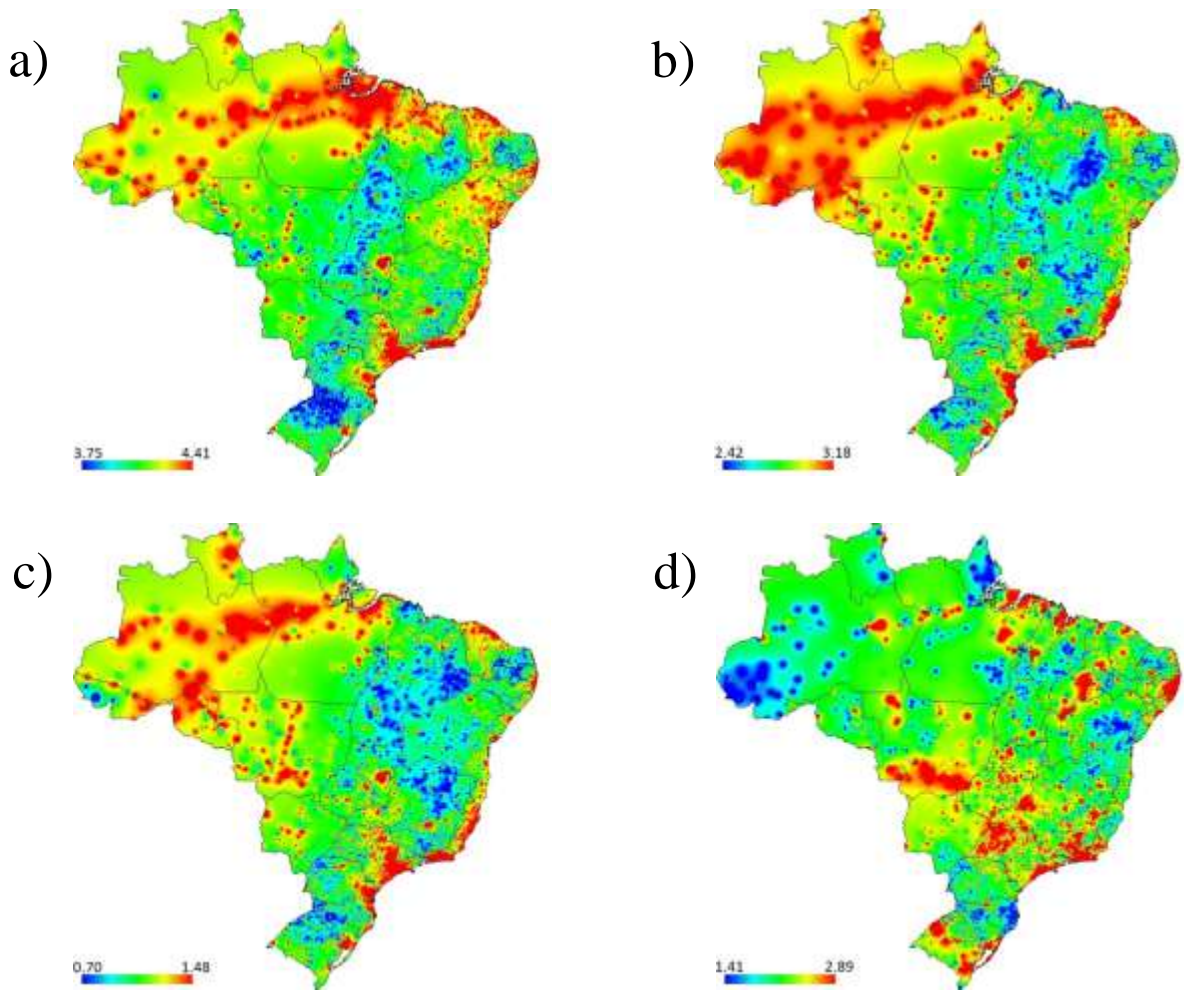

**Fig. 3.** a) population, b) number of confirmed cases, c) number of deaths, on logarithmic scale, and d) the death rate on linear scale, along the study period. The adopted color coding displays in blue values at or below the 1st quartile, and red for values at and above the 3rd quartile.

Time evolution of the percentage of population confirmed positive for coronavirus, deaths per 100000 inhabitants, and the death rate are presented in Figs. 4, 5 and 6, respectively. Cumulative values in twelve consecutive 30 day windows are used, from 4/4/2020 to 3/30/2021, so that month labels in the figures correspond to predominant months of the consecutive thirty-day periods. As these variables are all scaled in terms of municipality size, linear scales are used, with color coding adjusted in each case to emphasize the difference among the periods of observation and the geographic regions.

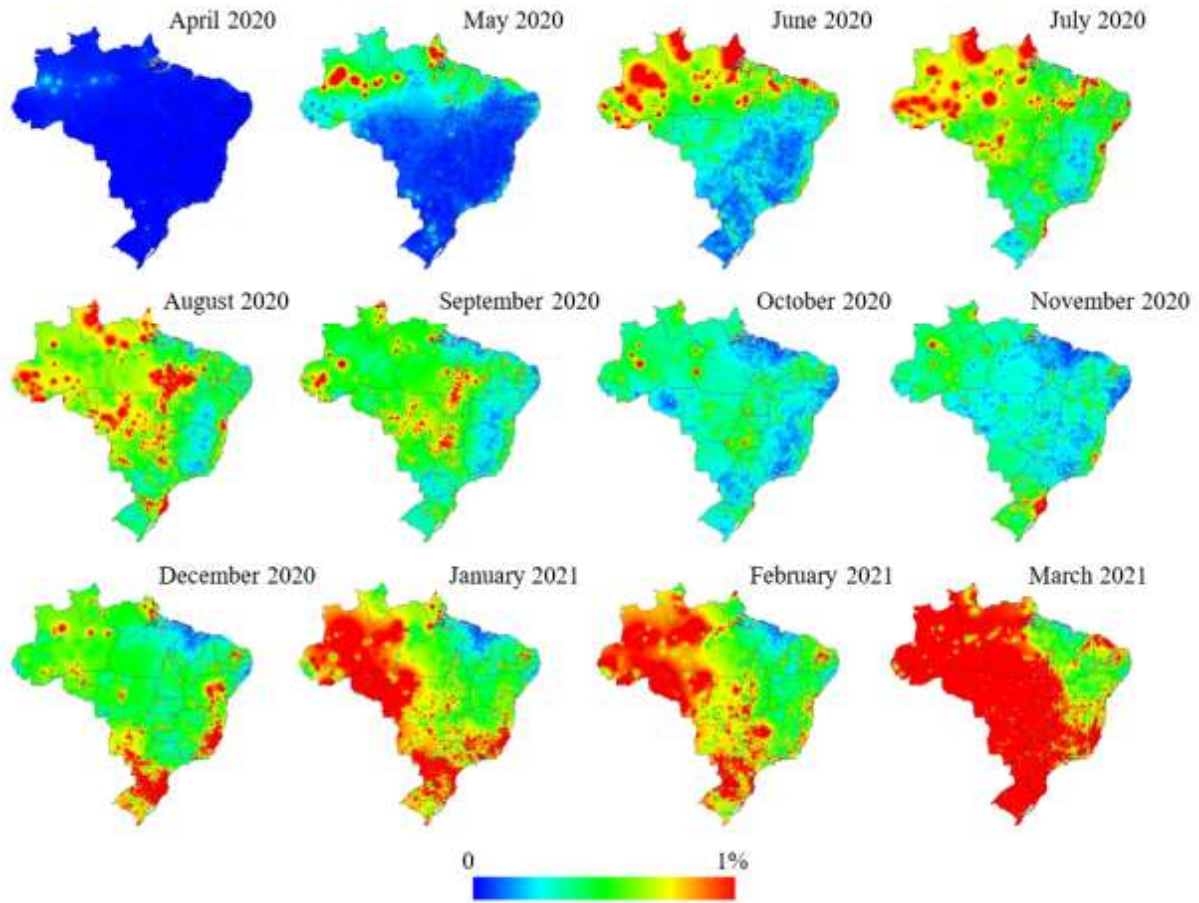

**Fig. 4.** Percentage of population confirmed positive for coronavirus in consecutive 30 day windows.

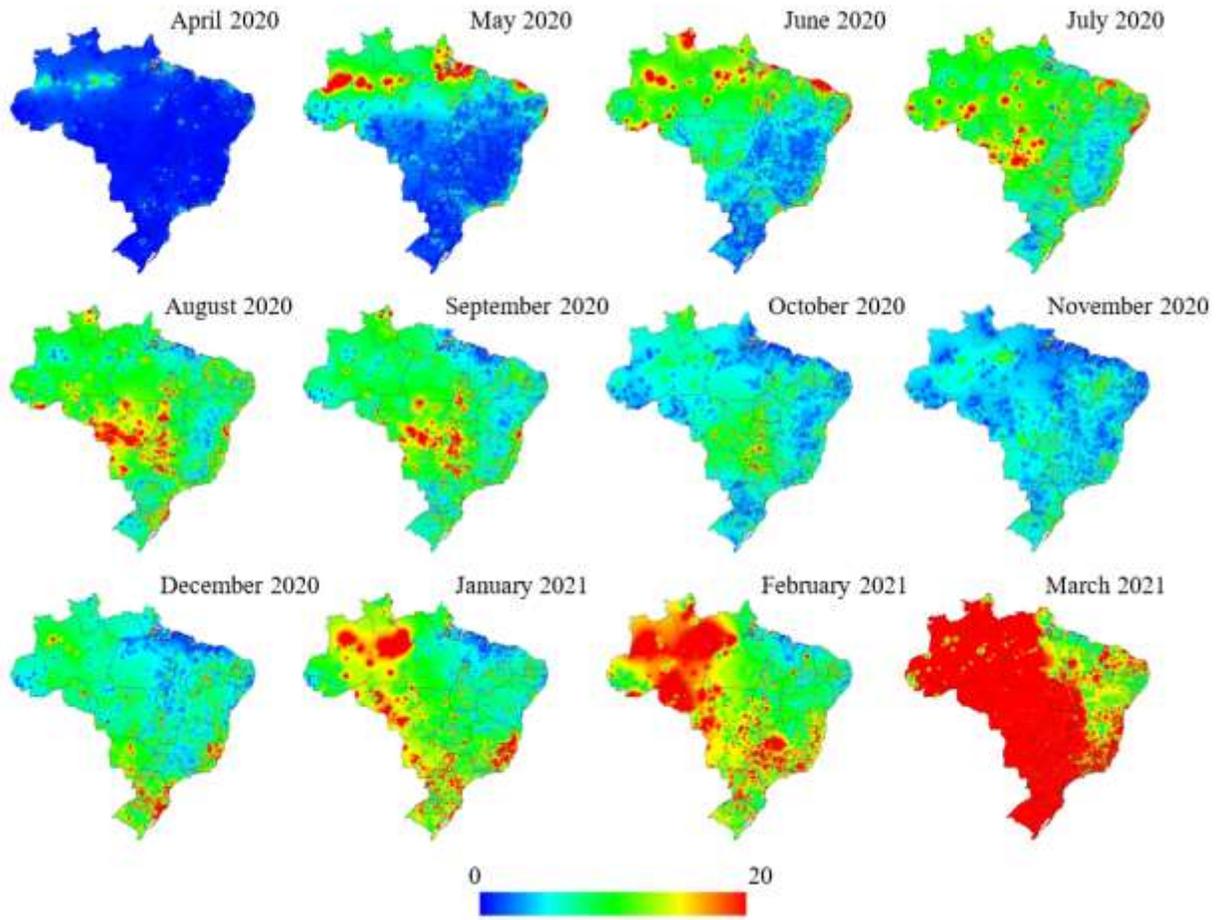

**Figure 5.** Number of deaths per 100000 inhabitants, in consecutive 30 day windows.

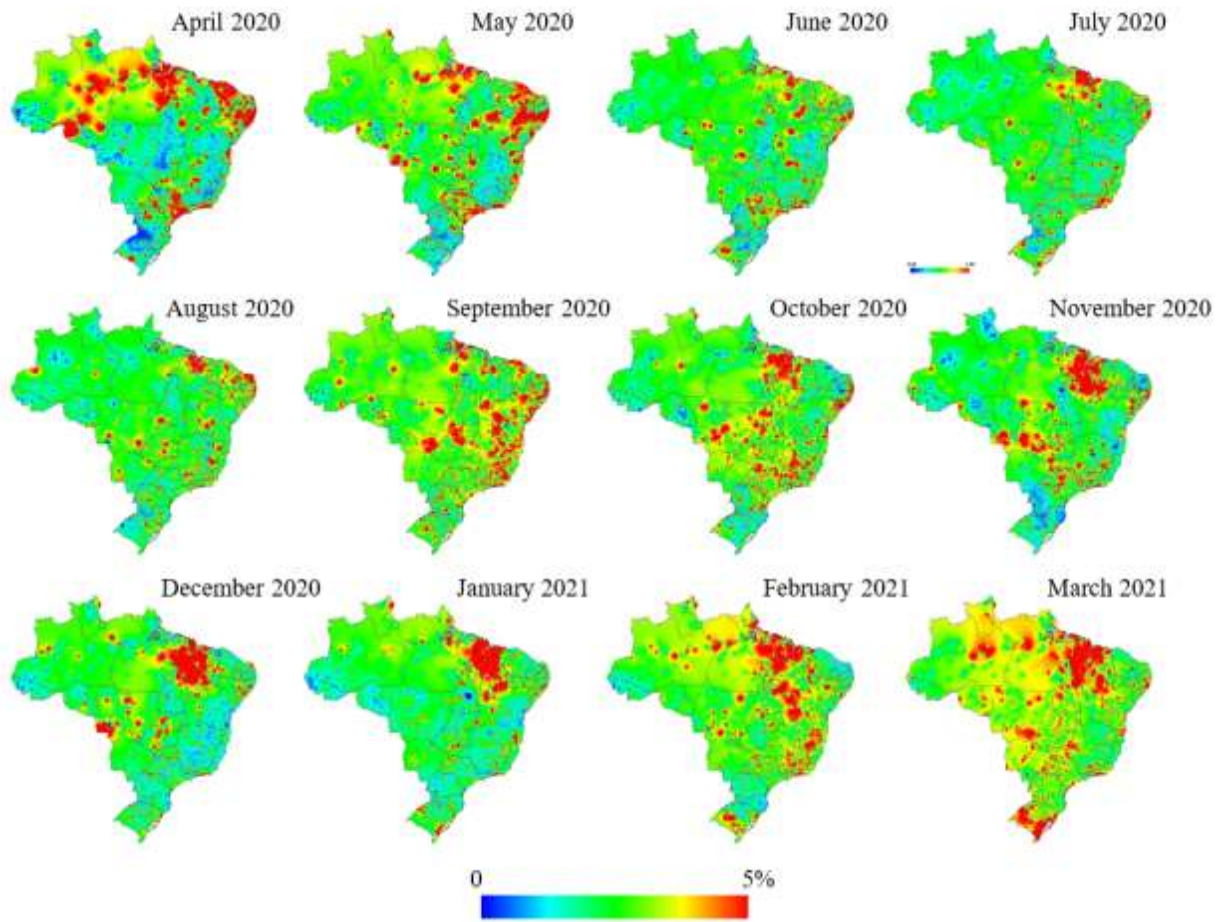

**Figure 6.** Percentage of confirmed cases that end up in deaths, in consecutive 30 day windows.

By comparing Figs. 4 and 5, it is seen that the number of deaths per 100K inhabitants rather closely follows the population confirmed cases percentage, both in terms of time evolution and spatial distribution, as perhaps may be expected. On the other hand, the death rate (number of deaths per number of confirmed cases) presented in Fig. 6 shows alarmingly high values in different regions of the country, at different timeframes, and does not seem to be related directly with the known peak in August 2020, and the ongoing peak since the beginning of 2021, in terms of confirmed cases and/or deaths. It follows that this variable should be scrutinized more closely in terms of association with individual mitigation measures taken locally by state and municipal governments.

The results of the DEA analysis, both in terms of the efficiency frontier, and the inefficiency frontier, for both constant and variable returns to scale, is shown in Fig. 7. While inefficiency ($I$) may be seen as a complement of efficiency ($E$) to unity (e.g. as $I = 1 - E$), it is important to emphasize here that the DEA method relies heavily on the concept of outliers. More precisely, if a single municipality exhibits zero deaths and/or confirmed cases in a given time frame, it automatically becomes part of the efficiency frontier in the input oriented DEA version, and distorts (reduces) the efficiency scores of all the other municipalities. The output oriented inefficiency frontier is on the other hand formed by municipalities with most confirmed cases and/or most deaths, and is therefore less affected by the outlier influence of small municipalities. This is the reason why the 128 very small municipalities with zero reported deaths have been excluded from the current study, out of the total of 5570.

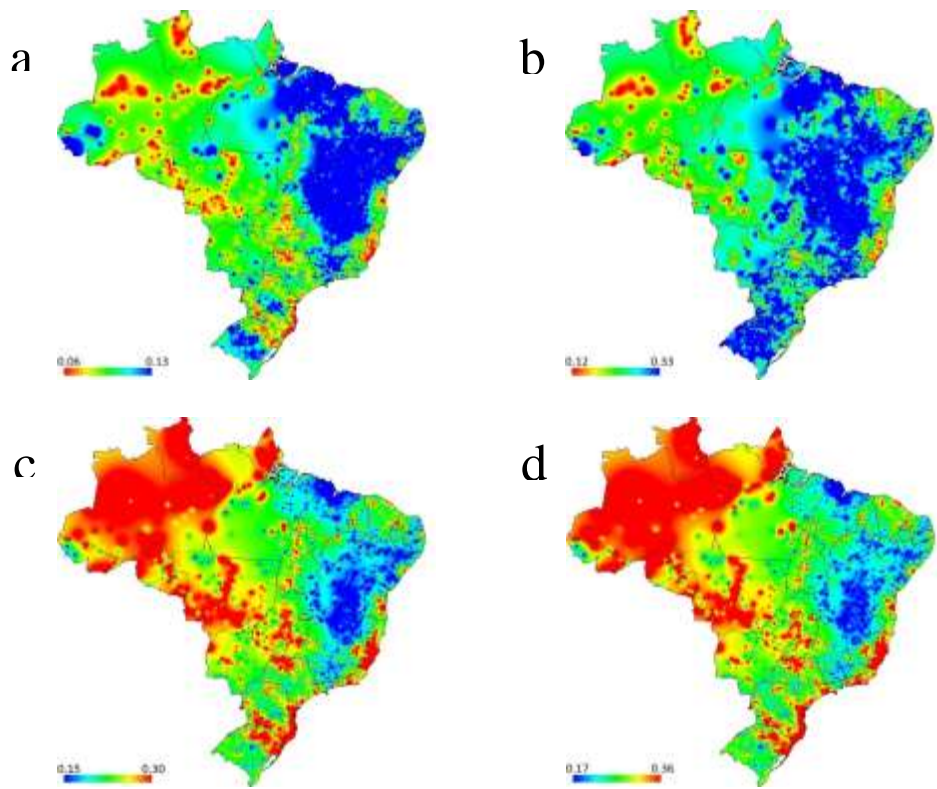

**Figure 7.** a) DEA CRS efficiency, b) DEA VRS efficiency, c) DEA CRS inefficiency, and d) DEA VRS inefficiency, along the study period. The adopted color coding displays in blue values at or below the 1st quartile, and red for values at and above the 3rd quartile for the efficiency calculations, and is inverted for inefficiency: red means "bad practice", and blue means "not so bad practice".

While the performance of Brazilian municipalities generally agrees between the two scale variants of efficiency (Figs. 7a and 7b) with those of inefficiency (Figs. 7c and 7d) frontiers, the emphasis in the former approach lies on highlighting small cities with few confirmed cases and deaths, and tends to underrepresent the cities with low efficiency. By comparing the CRS with the VRS versions (Fig. 7a with Fig. 7b, and Fig. 7c with Fig.7d) it can be concluded that taking into account scaling effect does not bring about a significant change of the spatial distribution of results, probably because the choice of input variables already accounts for scale in terms of population size, for the full study period.

As already mentioned, the efficiency frontier approach is rather sensitive to presence of outliers (small cities with zero confirmed cases and/or deaths), and it is found that it yields reasonable results only for the whole period of study, while the time evolution in 30-day windows enhances the outlier effect. The time evolution study is therefore implemented here only for the inefficiency frontier, with the results displayed in Figs 8 and 9.

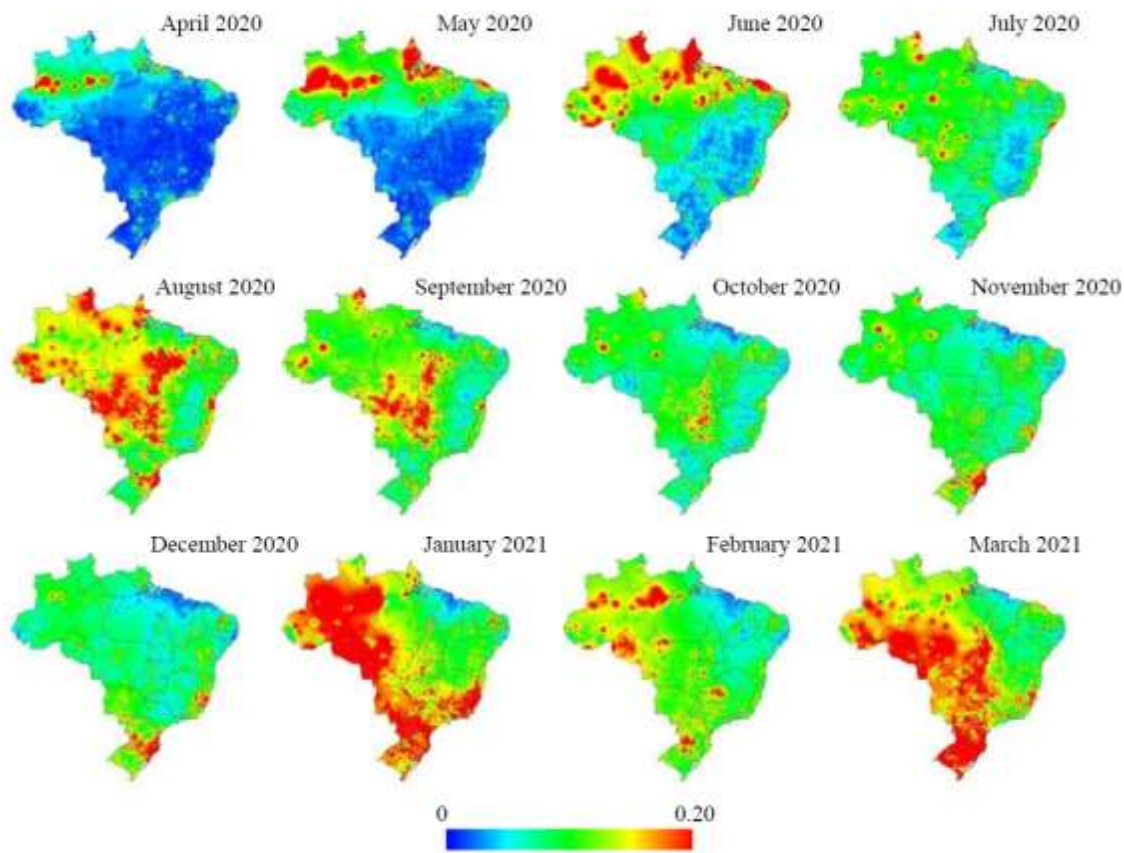

**Figure 8.** DEA CRS inefficiency, in consecutive 30 day windows.

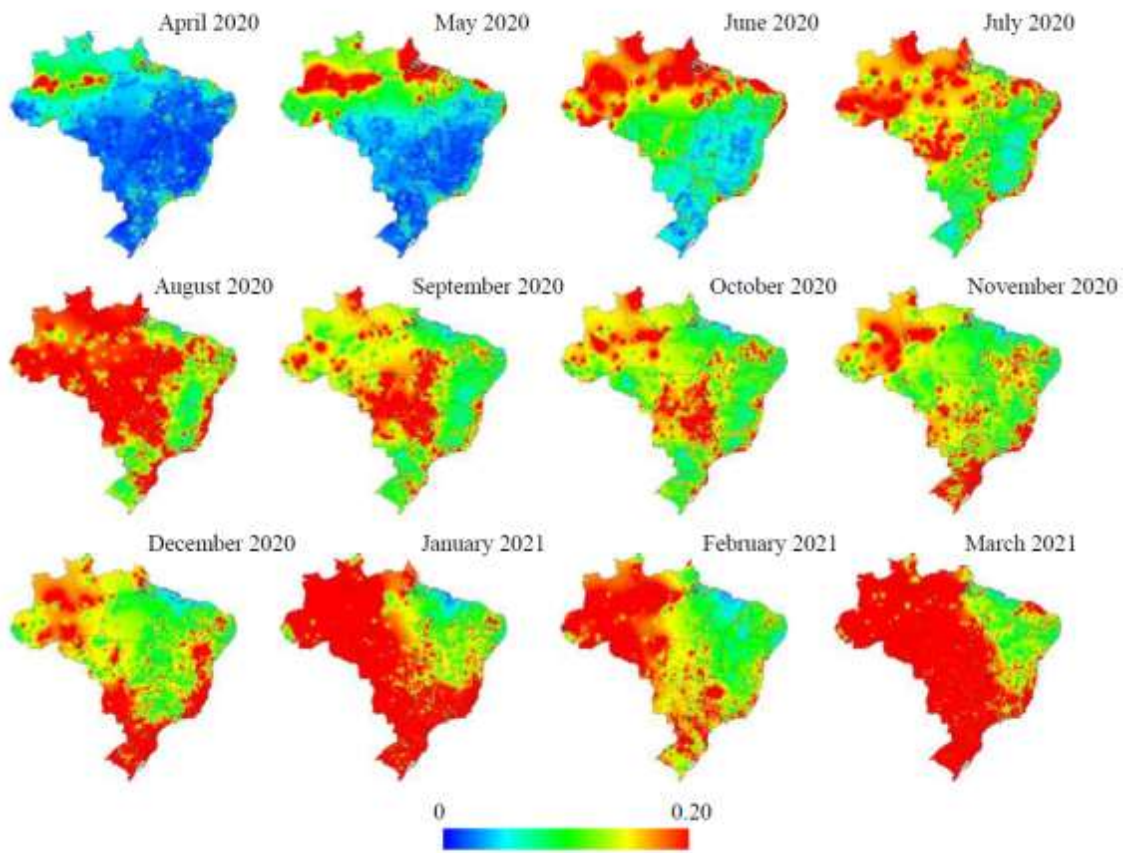

**Figure 9.** DEA VRS inefficiency, in consecutive 30 day windows.

Comparing the results for individual months on Figs. 8 and 9 it can be concluded that monthly cumulative values are not sufficient to account for scale, as the August 2020 peak and the ongoing peak (at the time of writing this article, March-April 2021) are better emphasized in the scale sensitive VRS approach. It should be noted here that both CRS and VRS versions demonstrate less areas in red (high inefficiency) in February 2021 in comparison with January 2021, but from Figs. 5 and 6 it is seen that both the number of deaths and the death rate have increased from January to February 2021 in the ongoing pandemic peak, while the confirmed cases number has not grown considerably, as can be seen in Fig. 4. This means that the inefficiency frontier has shifted forward, to municipalities with an increased number of deaths, but this is observed only in a subset of the municipalities that were the most inefficient in February 2021. The DEA results presented in Figs. 8 and 9 highlight the inefficiency scores relative to the most inefficient ones in each particular period of time, those that comprise the inefficiency frontier. Therefore, the authorities of the municipalities that show as red in February 2021 should be even more concerned than those of the municipalities shown as red for January, while in March 2021the situation becomes rather uniformly alerting among most municipalities in Brazil, perhaps less so for the northeast region.

**Conclusions**

The current study aims to provide a contribution to the assessment of the current observational data on the COVID-19 virus outbreak in Brazil, from a phenomenological point of view. The current analysis of the data reveals the spatial distribution of the hotspots in this continental size country over the entire period of the pandemic, as well as the evolution on monthly scale.

Full-fledged videos corresponding to Figs. 4, 5, 6, 8 and 9, comprised of 384 frames each (for a total of 413 observation days from February 25, 2020 to April 12,2021, there are 384 distinct 30-day overlapping timeframes) are available in the supplementary material [12]. Additional calculation results (for these, or for different timeframes, in the form of data tables, images and/or videos) are available from the author upon request, in the hope that such information may subsidize decision making through comparative analysis of previous practice, and thus contribute to pandemic mitigation efforts.

**Acknowledgments:**

**Funding:** The author acknowledges support of Brazilian agency CNPq through the research grant 307445/2018-6.

**Competing interests:** The author declares that he has no competing interests.

**Data and materials availability:** The data is available at (*1*). All the code was developed by the author in C programming language (with inline assembler code, for DEA calculations) and CUDA, C and Windows APIs (for interpolation and visualization). The source code, Visual Studio projects and executables are available from the author upon request.